\documentclass[%
 reprint,
superscriptaddress,
amsthm,amsmath,amssymb,
 aps,
]{revtex4-1}

\usepackage{graphicx}
\usepackage{dcolumn}
\usepackage{bm}
\usepackage{physics}
\usepackage{mathrsfs}
\usepackage{subfigure}

\begin{document}

\title{Realizing fully reference-frame-independent quantum key distribution by exploiting quantum discord}

\title{Realizing fully reference-frame-independent quantum key distribution by exploiting quantum discord}
\author{Rong Wang} 
\email{rwangphy@hku.hk}
\affiliation{Department of Physics, University of Hong Kong, Pokfulam Road, Hong Kong SAR, China}
\author{Chun-Mei Zhang}
\email{cmz@njupt.edu.cn}
\affiliation{Institute of Quantum Information and Technology, Nanjing University of Posts and Telecommunications, Nanjing, China}
\affiliation{State Key Laboratory of Cryptology, Beijing, China}






\begin{abstract}
Reference-frame-independent quantum key distribution was proposed to generate a string of secret keys without a shared reference frame. Based on the Bloch sphere, however, the security analysis in previous methods is only independent on azimuthal angle, while a reference frame is determined by both polar angle and azimuthal angle. Here, we propose a $3 \times 3$ matrix whose singular values are independent on both polar angle and azimuthal angle, as well as take advantage of quantum discord, to realize a fully reference-frame-independent quantum key distribution. Furthermore, we numerically show that the performance of our method can reduce to the previous one if the key generation basis is calibrated.
\end{abstract}


\pacs{Valid PACS appear here}
\maketitle


\section{\label{sec:level1}Introduction}
\begin{figure}[h]
	\centering
	\includegraphics[width=0.6\linewidth]{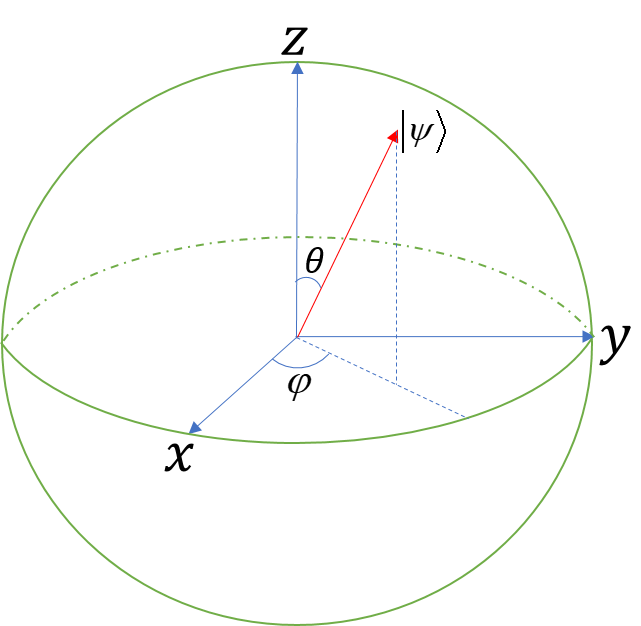}
	\caption{Characterization of a reference frame based on the Bloch sphere.}
	\label{bloch}
\end{figure}

Based on the principles of quantum physics, quantum key distribution (QKD) \cite{BB84} allows two legitimate parties, Alice and Bob, to exchange  information-theoretic secret keys, in the presence of an eavesdropper, Eve. Generally, a shared reference frame, which provides a vital benchmark for Alice and Bob to encode and decode quantum states, is essential to guarantee the regular running of practical QKD systems \cite{xavier2009experimental,tang2014experimental,ding2017polarization,wang2015experimental,tang2016measurement,yin2016measurement,boaron2018secure,comandar2016quantum}. As illustrated in Fig. \ref{bloch}, a reference frame can be characterized by both polar angle $\theta$ and azimuthal angle $\varphi$. To establish a common reference frame, it is indispensable to perform some real-time calibration, e.g., the alignment of polarization states for polarization coding or the compensation of phase difference for phase coding, which, however, decreases the generation of secret keys and complicates the operation of QKD systems. 

Fortunately, reference-frame-independent QKD (RFI-QKD) \cite{laing2010reference,zhu2022improved} can generate secret keys with an unknown and slowly drifted reference frame. However, there is a stringent assumption that the key-generation basis, denoted as the $\sigma_z$ basis, should be calibrated well, that is, $\theta=0$. With the premise of a well aligned $\sigma_z$  basis, RFI-QKD \cite{laing2010reference,zhu2022improved} is robust against the unknown and slow drift of the test bases, denoted as the $\sigma_x$  and $\sigma_y$  bases, that is, $\varphi$ may change with time slowly. Nevertheless, it is tricky to calibrate Alice and Bob's key-generation basis accurately. For example, in the polarization coding scheme, all polarization states may be rotated with the transmission in the optical fiber \cite{xavier2009experimental,tang2014experimental,ding2017polarization}, which violates the assumption that the key-generation basis is aligned well. Therefore, a natural question is, can we design a QKD protocol which is independent on both polar angle $\theta$ and azimuthal angle $\varphi$?

Here, we propose a fully reference-frame-independent QKD (FRFI-QKD) protocol to solve this question. Inspired by viewing quantum discord \cite{henderson2001classical,ollivier2001quantum,modi2012classical} as a resource and its application in QKD \cite{pirandola2014quantum,wang2023quantum,wang2023secure}, we construct a $3 \times 3$ matrix whose singular values are independent on both $\theta $ and $\varphi $, and exploit quantum discord to realize a FRFI-QKD protocol, which is robust against the slow drift of both $\theta $ and $\varphi $. Furthermore, we numerically demonstrate that the performance of the proposed FRFI-QKD protocol can reduce to the previous RFI-QKD protocol \cite{laing2010reference,zhu2022improved} if the key-generation basis is well calibrated.


\section{PROTOCOL DESCRIPTION}
The protocol of FRFI-QKD runs as follows, which is similar to RFI-QKD \cite{laing2010reference,zhu2022improved} .\\
	$ \bullet $ \textbf{State preparation:} Alice randomly chooses a basis ${\xi _A}$ from her local basis set $\{ {\sigma _x},{\sigma _y},{\sigma _z}\} $ and a bit $b \in \{0, 1\}$ to prepare the single-photon quantum state, and sends it to Bob through the insecure quantum channel.\\
	$ \bullet $ \textbf{Measurement:}  Bob randomly chooses a basis ${\xi _B}$ from his local basis set $\{ {\sigma _x},{\sigma _y},{\sigma _z}\} $ to measure the received quantum state, and records the corresponding results.\\
	$ \bullet $ \textbf{Classification:}  Alice publicly announces her basis for each preparation, and Bob publicly announces his basis for each measurement. According to their basis information, Alice and Bob classify their events into groups ${G_{{\xi _A}{\xi _B}}}$, where ${{\xi _A}{\xi _B}}$ denotes different combinations of their bases.\\
	$ \bullet $ \textbf{Parameter estimation:} Alice and Bob randomly sample a part of their raw data to estimate the gains and quantum bit error rates.\\
	$ \bullet $ \textbf{Post-processing:} Alice and Bob perform the reverse information reconciliation and privacy amplification for the raw data on $\sigma_z \otimes \sigma_z$ to obtain the final secret key. 

\section{security analysis}

This section shows the security proof. Let $\rho_{AB}$ be two-qubit state shared by Alice and Bob, in Pauli representation, it can be expressed by 
\begin{equation}
\begin{aligned}
\rho_{AB}=&\frac{1}{4}(\mathbb{I} \otimes \mathbb{I} +\sum_{i} a_i \sigma_i \otimes \mathbb{I} + \sum_{j} b_j \mathbb{I} \otimes \sigma_j \\
                 &+\sum_{i,j} T_{ij} \sigma_i \otimes \sigma_j),\\
\end{aligned}
\end{equation}
where $\mathbb{I}$ is the identity matrix, $ \sigma_i$ and $\sigma_j$ are the three Pauli matrices with $i,j \in \{x, y, z\}$, $a_i=\Tr[\rho_{AB} \sigma_i \otimes \mathbb{I}]$ and $b_j=\Tr[\rho_{AB} \mathbb{I} \otimes \sigma_j]$ are components of the local Bloch vectors, and $T_{ij}=\Tr[\rho_{AB} \sigma_i \otimes \sigma_j]$ are components of the $3 \times 3$ correlation tensor matrix. In the following, similar to previous works \cite{renner2008security,acin2007device,pironio2009device}, we reduce $\rho_{AB}$ to a Bell diagonal one without limiting Eve's power and changing the observable statistics. It has been proved that \cite{horodecki1996perfect,horodecki1996information}, under some appropriate local unitary transformation $U_A \otimes U_B$ where $U_{A,B} \in SU(2)$, we have 
\begin{equation}
\begin{aligned}
\rho'_{AB}:=U_A \otimes U_B \rho_{AB} U^{\dagger}_A \otimes U^{\dagger}_B
\end{aligned}
\end{equation}
with a diagonal correlation tensor matrix. In other words, if denoted its components by $T'_{ij}$, we have $T'_{i \neq j}=0$. Moreover, we can assume without loss of generality that $T'_{zz} \ge T'_{xx} \ge |T'_{yy}| \ge 0$ (see lemma 6 in \cite{wang2023quantum}). Then we introduce 
\begin{equation}
\bar{\rho}_{AB}:=\frac{1}{4}(\rho'_{AB}+\sum_{i} \sigma_i \otimes \sigma_i \rho'_{AB} \sigma_i \otimes \sigma_i),
\end{equation}
whose correlation tensor matrix is totally same as that of $\rho'_{AB}$, while the components of the local Bloch vectors are vanishing. And it is straightforward to see that $\bar{\rho}_{AB}$ is Bell diagonal. 
By operating the inverse transformation of $U_A \otimes U_B$ on $\bar{\rho}_{AB}$, we obtain
\begin{equation}
\begin{aligned}
\sigma_{AB}:= & U^{\dagger}_A \otimes U^{\dagger}_B \bar{\rho}_{AB} U_A \otimes U_B \\
                     = & \frac{1}{4}(\mathbb{I} \otimes \mathbb{I} + \sum_{i,j} T_{ij} \sigma_i \otimes \sigma_j), \\
\end{aligned}
\end{equation}
whose correlation tensor matrix is totally same as that of $\rho_{AB}$. Comparing $\sigma_{AB}$ and $\rho_{AB}$, they produce same observable statistics while $\sigma_{AB}$ is more mixed than $\rho_{AB}$. Therefore, Eve can gain at least the same knowledge or even more, if Alice and Bob share $\sigma_{AB}$ \cite{renner2008security,acin2007device,pironio2009device}. Since $\sigma_{AB}$ is locally equivalent to a Bell diagonal sate, $\bar{\rho}_{AB}$, under $U_A \otimes U_B$, we can write it as
\begin{equation}
\begin{aligned}
U_A \otimes U_B \sigma_{AB} U^{\dagger}_A \otimes U^{\dagger}_B= \sum_{k=1}^4 \lambda_k \ket{\Phi_k}\bra{\Phi_k}, 
\end{aligned}
\end{equation}
where $\ket{\Phi_{1,2}}=(\ket{00}\pm \ket{11})/\sqrt{2}$, $\ket{\Phi_{3,4}}=(\ket{01}\pm \ket{10})/\sqrt{2}$. The parameters have following relations that (see lemma 6 in \cite{wang2023quantum})
\begin{equation}
\label{parametersrelation}
 \left\{
\begin{array}{lc}
\lambda_1 + \lambda_2 + \lambda_3 + \lambda_4 = 1 \\
\lambda_1 + \lambda_2 - \lambda_3 - \lambda_4 = T'_{zz}\\
\lambda_1 - \lambda_2 + \lambda_3 - \lambda_4 = T'_{xx}\\
\lambda_1 - \lambda_2 - \lambda_3 + \lambda_4 =-T'_{yy}\\
\lambda_1 \ge \lambda_2 \ge \lambda_3 \ge \lambda_4 \ge 0\\
\end{array}.
\right.
\end{equation}
According to the previous results \cite{datta2008studies,luo2008quantum,lang2010quantum}, we immediately have (see lemma 7 and corollary 8 in \cite{wang2023quantum} also)
\begin{equation}
\label{conditionalentropy}
\begin{aligned}
H(Z_B|E) \ge D(A|B) = 1&-(\lambda_1 + \lambda_2)h(\frac{\lambda_2}{\lambda_1 + \lambda_2}) \\
                                       &-(\lambda_3 + \lambda_4)h(\frac{\lambda_4}{\lambda_3 + \lambda_4}), \\
\end{aligned}
\end{equation}
so that we can obtain the key rate with the parameters $\lambda_k$ derived from the observable statistics. To do so, we focus on the observable correlation matrix in the following
\begin{equation}
\begin{aligned}
\hat{T}:=
\begin{pmatrix}
\left \langle \sigma_x \otimes \sigma_x \right \rangle & \left \langle \sigma_x \otimes \sigma_y \right \rangle & \left \langle \sigma_x \otimes \sigma_z \right \rangle
 \\
\left \langle \sigma_y \otimes \sigma_x \right \rangle & \left \langle \sigma_y \otimes \sigma_y \right \rangle & \left \langle \sigma_y \otimes \sigma_z \right \rangle
 \\
\left \langle \sigma_z \otimes \sigma_x \right \rangle & \left \langle \sigma_z \otimes \sigma_y \right \rangle & \left \langle \sigma_z \otimes \sigma_z \right \rangle
\end{pmatrix},
\end{aligned}
\end{equation}
where $\left \langle \sigma_i \otimes \sigma_j \right \rangle =T_{ij}$. 
Considering the fact that there is always a unique rotation $O \in SO(3)$ corresponding to any unitary transformation $U \in SU(2)$, there exists $O_{A,B} \in SO(3)$ so that $O_A \hat{T} O_B^t =$ diag$[T'_{zz}, T'_{xx}, T'_{yy} ]$, where diag$[T'_{zz}, T'_{xx}, T'_{yy} ]$ is the diagonal correlation matrix of $\bar{\rho}_{AB}$.
For simplicity, we could also use its singular value decomposition and the singular values are the eigenvalues of $\sqrt{\hat{T}^t\hat{T}}$. Denoting the singular values with decreasing order, that is, $\{T_1, T_2, T_3\}$, we have $T_1=T'_{zz}$, $T_2=T'_{xx}$ and $T_3=|T'_{yy}|$. In the asymptotic scenario, with the relations Eq. \eqref{parametersrelation}, Eq. \eqref{conditionalentropy}, above singular values and Devetak-Winter bound \cite{devetak2005distillation}, the secret key rate is given by 
\begin{equation}
\label{keylength}
\begin{aligned}
R   & \ge 1-h(Q)-\frac{1+T_1}{2}h(\frac{1+T_1-T_2-T_3}{2(1+T_1)}) \\
     & -\frac{1-T_1}{2}h(\frac{1-T_1-T_2+T_3}{2(1-T_1)}), \\
\end{aligned}
\end{equation}
where $Q:=(1- \left \langle \sigma_z \otimes \sigma_z \right \rangle)/2$ is actually the bit error rate. 


\section{Simulation}
In the simulation, we assume the dark count rate of single-photon detectors is $10^{-6}$, and the misalignment error rate is $1.5\%$. To illustrate the robustness against the variation of reference frame, we investigate the secret key rate of FRFI-QKD under different $\theta$ and $\varphi $ with respect to the overall loss (including the detection efficiency of single-photon detectors) between Alice and Bob. Note that $\theta$ and $\varphi $ are actually unknown to Alice and Bob, which are set to be known only for ease of simulation. We also compare the performance of FRFI-QKD with RFI-QKD \cite{laing2010reference,zhu2022improved} and six-state QKD \cite{bruss1998optimal}. 

\begin{figure}[h]
\centering
\includegraphics[width=\linewidth]{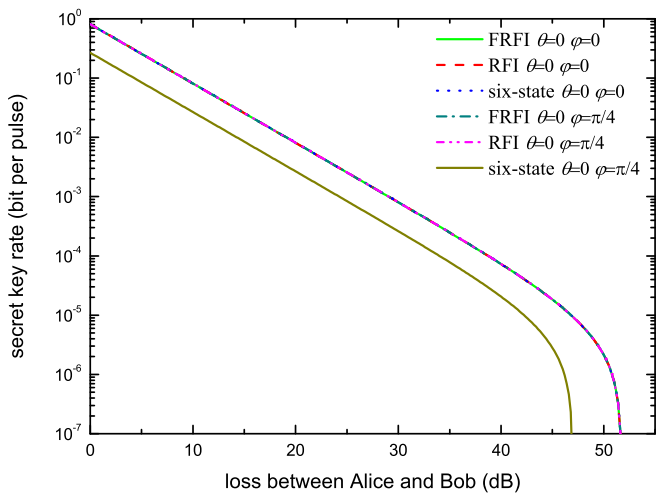}
\caption{Secret key rate (per pulse) as a function of the overall loss between Alice and Bob, where $\theta  = 0$ and $\varphi=0,\pi /4$. Except for the result of six-state QKD under $\theta  = 0$ and $\varphi  = \pi /4$ which is shown in the bottom curve, the results of other cases completely overlap on the top curve.}
\label{theta0-faiX}
\end{figure}

Fig. \ref{theta0-faiX} illustrates the results of FRFI-QKD, RFI-QKD and six-state QKD under $\theta  = 0$ and $\varphi  = 0,\pi /4$. Here, we assume the polar angle $\theta$ of Alice and Bob's reference frame is the same, that is, their key-generation basis is well aligned, while the relative azimuthal angle $\varphi$ of their reference frame is not aligned. It clearly demonstrates that, in the case of well-aligned key-generation basis, FRFI-QKD and FRI-QKD shows the same performance, and their performance is irrelevant to the variation of the azimuthal angle $\varphi$; on the contrary, the performance of six-state RFI-QKD is severely depressed by the relative variation of $\varphi$.

\begin{figure}[h]
\centering
\includegraphics[width=\linewidth]{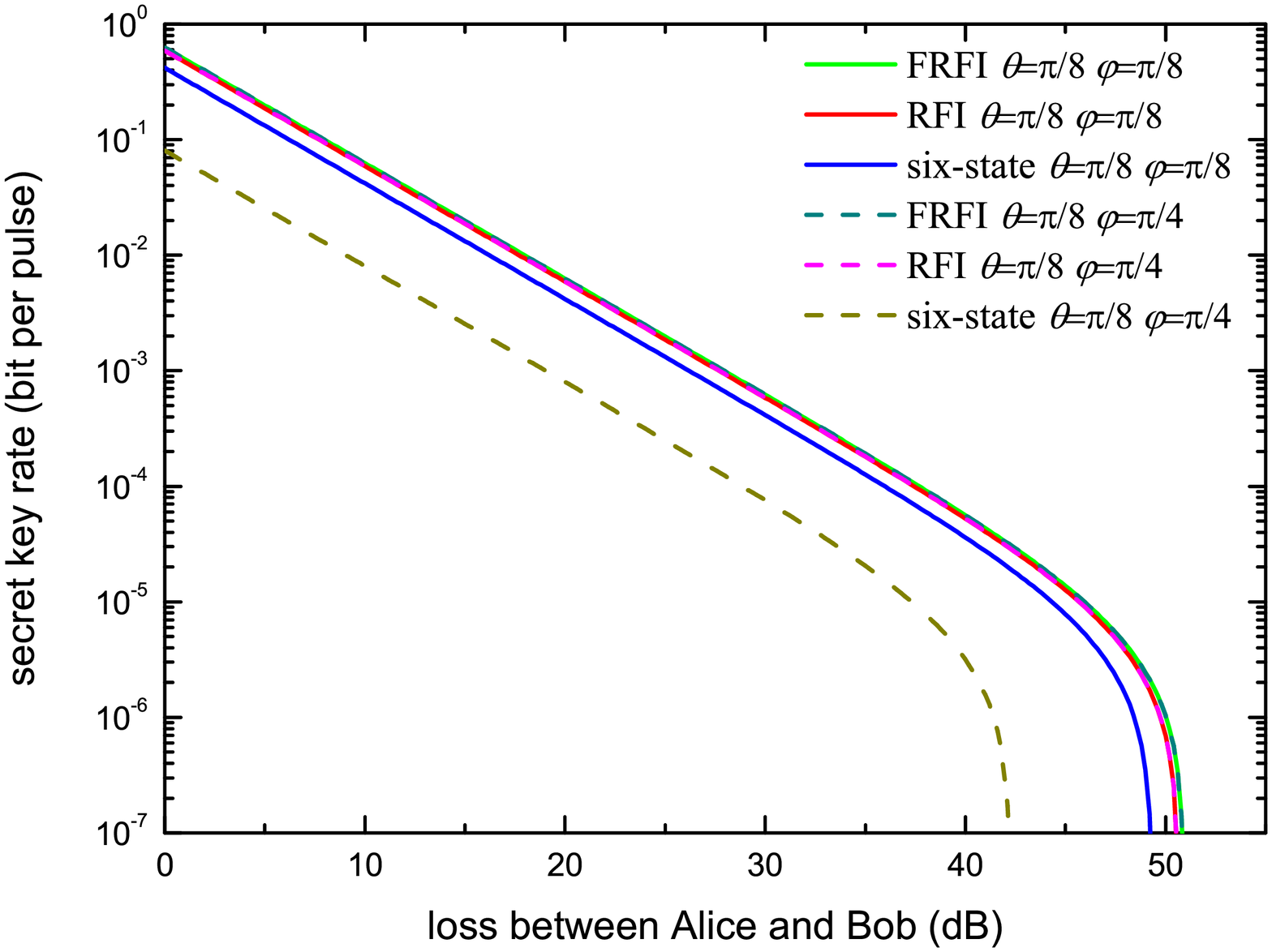}
\caption{Secret key rate (per pulse) as a function of the overall loss between Alice and Bob, where $\theta  = \pi /8$ and $\varphi  = \pi/8,\pi /4$.}
\label{theta0.125pi-faiX}
\end{figure}

Fig. \ref{theta0.125pi-faiX} illustrates the results of FRFI-QKD, RFI-QKD and six-state QKD under $\theta  = \pi/8$ and $\varphi  =\pi/8,\pi /4$. Compared with Fig. \ref{theta0-faiX}, it can be seen that a small misalignment on the key generation basis affects the performance of these three protocols, and FRFI-QKD the most slightly. On the other hand, with the premise of $\theta  = \pi/8$, the performance of FRFI-QKD and RFI-QKD is invariant with the drifting of the azimuthal angle $\varphi$, and the performance of six-state RFI-QKD is depressed by $\varphi$. 

\begin{figure}[h]
	\centering
	\includegraphics[width=\linewidth]{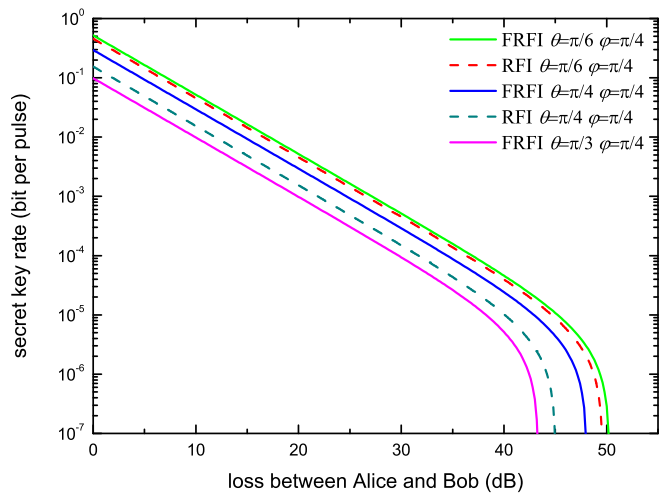}
	\caption{Secret key rate (per pulse) as a function of the overall loss between Alice and Bob, where $\theta  = \pi /6, \pi/4, \pi/3$ and $\varphi  = \pi /4$.}
	\label{thetaX-fai0.25pi}
\end{figure}

Fig. \ref{thetaX-fai0.25pi} illustrates the comparison results of FRFI-QKD and RFI-QKD under $\theta  = \pi /6, \pi/4, \pi/3$ and $\varphi  = \pi /4$ where six-state QKD cannot produce any key. With the increase of $\theta$, FRFI-QKD exhibits more robustness than RFI-QKD. In particular, when the key generation basis has a large misalignment (e.g. $\theta=\pi/3$), FRFI-QKD can generate considerable key rate while RFI-QKD cannot, which demonstrates that the real-time calibration requirement of a reference frame can be greatly relaxed with our FRFI-QKD protocol.

\section{Discussion and Conclusion}
In summary, based on the proposed $\hat{T}$-matrix and using quantum discord, we have proposed the FRFI-QKD protocol and analyzed its security against collective attacks, which is robust against the slow drift of both polar angle $\theta$ and azimuthal angle $\varphi$. Simulation results show that, the performance of FRFI-QKD is totally independent on the azimuthal angle $\varphi$ of the reference frame, and slightly dependent on the polar angle $\theta$ of the reference frame. Interestingly, even when the key-generation basis suffers from a large rotation, our protocol can produce considerable key rate, which can greatly relax the real-time calibration requirement of a reference frame, and find its applications in more complicated reference frame scenarios. Moreover, when the key-generation basis is well aligned, our FRFI-QKD protocol exhibits the same performance as RFI-QKD. At last, to combat the potential attack aimed at the measurement device, we can naturally extend our protocol into the measurement-device-independent scenario \cite{braunstein2012side,lo2012measurement}.

It should be noted that while tomography-based QKD \cite{watanabe2008tomography,zhan2020tomography} can generate an almost positive key rate without knowledge of polar and azimuthal angles, its security \cite{watanabe2008tomography,zhan2020tomography} depends heavily on the assumption that Eve's attack can be modeled as a complete positive trace preserving (CPTP) map. This security is only guaranteed if Bob is required to output a classical qubit even in the absence of a quantum signal, leading the loss-intolerant feature of tomography-based QKD.

\begin{acknowledgments}
We thank Hoi-Kwong Lo for helpful discussions. R. W is supported by the University of Hong Kong start-up grant, Chun-Mei Zhang is supported by China Postdoctoral Science Foundation (2019T120446, 2018M642281), Jiangsu Planned Projects for Postdoctoral Research Funds (2018K185C), the Natural Science Foundation of NJUPT (NY221058, 1311).
\end{acknowledgments}

\twocolumngrid

\nocite{*}

\bibliography{apssamp}

\end{document}